\documentclass[10pt]{article}

\usepackage{amsmath}
\usepackage{amsthm}

\usepackage[lined,boxed,commentsnumbered]{algorithm2e}

\newtheorem{definition}{Definition}

\newtheorem{remark}{Remark}

\newcommand{\mket}[1]{| #1 \rangle}

\newcommand{\nH}{\mathcal{H}}

\title{Twopartite, combinatorial approach to partial k-separability problem for general multipartite states}

\author{Roman Gielerak$^{(a)}$ and Marek Sawerwain$^{(a)}$ \\
(a) Institute of Control \& Computation Engineering, University of Zielona G\'ora \\
ul. Podg\'orna 50, 65-256 Zielona G\'ora, Poland,  \\
e-mail: {\it R.Gielerak@issi.uz.zgora.pl},  {\it M.Sawerwain@issi.uz.zgora.pl} \\
}

\begin{document}

\maketitle

\abstract{We describe a general methods to localize any sort of k-separability and therefore also the corresponding partial entanglement in genuinely multipartite mixed quantum states. Our methods are based exclusively on the known twopartite methods and some systematic procedures of combinatorial nature. Our methods are formalized in an algorithmic-like fashion and therefore they are easily implementable in a computer environments and might be effectively used for studying numerically these questions for sufficiently small systems}


\section{Introduction}

Quantum entanglement, since its discovery by Einstein, Podolsky and Rosen \cite{Key1}, and Schrodinger \cite{Key2} attracted very large attention, especially in the past two decades due to its intriguing properties and extremely important significance in the quantum information processing tasks \cite{Key3}, \cite{Key4}, \cite{Key11}. The main efforts  have been focused on bipartite entanglement and at present many interesting and deep results have been obtained in this case, see \cite{Key5}, \cite{Key6} for some recent reviews.

The present note is addressed to a very preliminary discussion of certain aspects of multipartite entanglement. The importance of multipartite entanglement can be illustrated by several illuminating examples. A class of the so called graph states used in one-way quantum computation model \cite{Key7} and fault-tolerant topological quantum computation \cite{Key21} model as well are both good examples. Multipartite entangled photons are essential resources used in quantum key distribution protocols \cite{Key8}, \cite{Key9}. For the purposes of quantum communication a multipartite entangled states can serve as quantum communication channels in most of the known teleportation protocols \cite{Key10}. Multipartite entanglement displays one of the most fascinating features of quantum theory that is called as nonlocality of the quantum world \cite{Key13}.

Comparing to the best known two-partite case \cite{Key4}, \cite{Key5}, \cite{Key6}, the situation concerning a general multipartite system is much more complicated even on the level of classification of the possible entanglements that can arise in such systems. Given a multipartite state, a natural question to ask is whether it is entangled state and if so how to measure the amount of entanglement included.

The main aim of the present note is to demonstrate how one can use exclusively two-partite systems methods to answer the first question for a general multipartite quantum state, see also \cite{Key18}, \cite{Key19} for some recent references on this topics. The presented in this note methods are being implemented and applied to study several particular (low dimensional mainly) systems within constructed in Zielona G\'ora University in the past few years simulator of quantum computation machine, called Quantum Computing System (QCS) which is presented partially in \cite{Key20} and will be presented in \cite{Key14} with all details necessary for this. The report from our computer experiments will be presented elsewhere.

The present note is organized in the following way. In the next section basic definitions will be introduced and the algorithm detecting and localizing entanglements in the case of pure states will be formulated. An extension of this method to general multipartite mixed states will be presented in Section~(\ref{lbl:general:multipartite:system:mixed:states:case}).

\section{A general multipartite systems. The case of pure states}

Let $Q$ be a quantum system composed from $n$ smaller and separated subsystems. The corresponding Hilbert spaces $\nH_i$ of the subsystems are assumed to be finite dimensional with the corresponding dimensions $k_i$ for $i=1, \ldots, n$.

Then the resulting Hilbert space $\nH$ describing possible states of the total system is equal to $\nH = \nH_1 \otimes \nH_2 \otimes \ldots \nH_n$ and has dimension $K = k_1 \cdot k_2 \cdot \ldots k_n$. The space of all quantum states i.e. the space of all nonnegative, of trace equal to one endomorphisms of the corresponding Hilbert space of states $\nH$ will be denoted by $E(\nH)$ and its boundary $\partial E(\nH)$ is composed from pure states of the system $Q$.

Let $\wp(n)$ stands for the set of all partitions of the set $I_n = \{1,2,\ldots,n\}$. For a particular $\pi = (X_1, \ldots, X_k) \in \wp(n)$ we denote the number of elements of $\pi$, $| \pi |=k$. A partial semiorder $\prec_f$ in $\wp(n)$ is introduced by the following definition: we say that $\pi = (X_1, X_2, \ldots, X_k)$ is finer than a partition $\pi^{'} = (X^{'}_1, X^{'}_2, \ldots, X^{'}_l)$, denoted as $\pi \prec_f \pi^{'}$ iff  for any $i \in \{1,2,\ldots, k\}$ there exists $j \in \{1,2,\ldots,l\}$  such that $X_i \subseteq X^{'}_{j}$. The maximal element $\pi_{\max}$ exists in the poset $(\wp(n),\prec_{f})$  and $\pi_{\max} = (I_n)$ and the corresponding minimal partitions $\pi_{\max} = (\{ 1 \},\{ 2 \}, \ldots, \{ n \})$.

The cardinality of the set $\wp(n)$ denoted as $|\wp(n)|$ is given by the so called Bell number $B(n)$. The number of all partitions of the set $I_n$ of length $k$ is given by Stirling number of second kind $S(n,k)$. In particular $S(n,2)=2^{n-1}-1$ and $B(n) = \sum^{n}_{k=1} S(n,k)$. 

With these preparatory definitions we can pass to the basic notions.

A pure state $\mket{\psi} \in \partial E(\nH_1 \otimes \nH_2 \otimes \ldots \otimes \nH_n)$, $n \geq 2$ is called $\pi$-separable iff 
\begin{itemize}
\item $\mket{\psi} = \bigotimes^{k}_{i=1} \mket{\psi_i}$ for $\pi = (X_1, X_2, \ldots, X_k) \in \wp(n)$, $\mket{\psi_i} \in \nH(X_i)$ and where $\nH(X_i)$ stands for the Hilbert space corresponding to the $X_i$-piece of the whole system, 
\item there is no finer partition $\pi^{'}$ for which $\mket{\psi}$  is $\pi^{'}$-separable.
\end{itemize}

\begin{remark}
There is some abuse of notation introduced in the above definition which however can be easily overcomed when passing to the corresponding matrix representation of $\mket{\psi}$.
\end{remark}

In particular a state $\mket{\psi}$ is called completely separable iff $\mket{\psi}$ is $\pi_{\min}$-separable. Any state $\mket{\psi}$ which is not $\pi_{\min}$-separable is partially entangled state as the following definition state.

\begin{definition}
A state $\mket{\psi} \in \partial E(\nH_1 \otimes \nH_2 \otimes \ldots \otimes \nH_n)$  is caled $\pi$-entangled state iff there exists $\pi \in \wp(n)$, $\pi \neq \pi_{\min}$ such that $\mket{\psi}$  is $\pi$-separable. In particular a state $\mket{\psi}$  which is $\pi_{\max}$-separable will be called completely entangled state.
\end{definition}

The first question when dealing with a particular state $\mket{\psi}$  is the question whether this state is separable or entangled state.

In the case of two-partite systems and for vector states the complete answer to this question is provided by the corresponding Schmidt decomposition \cite{Key3}, \cite{Key4}. However, in general n-partite case there is no canonical notion of Schmidt decomposition as is well known.

The algorithm formulated below gives a systematic way to answer this question in general and additionally localises the finest partition for which the separability holds.

To start with, we introduce the following construction, which can be easily implemented in any computer system which contains any of the standard Linear Algebra Package.

The canonical Schmidt decomposition: Let $\nH_1$, $\nH_1$ be two Hilbert spaces of dimension $n_1$, resp. $n_2$ . Then for any $\mket{\psi} \in \nH = \nH_1 \otimes \nH_2$, there exists an unique decomposition:
\begin{equation}
\mket{\psi} = \sum^{r(\mket{\psi|})}_{i=1} \lambda_i \mket{\psi^1_i} \otimes \mket{\psi^2_i}
\end{equation}
where the number $1 \leq r(\mket{\psi}) \leq \min(n_1, n_2)$ is the Schmidt rank of $\mket{\psi}$ and $\{ \mket{\psi^{\alpha}_{i}}, i=1,2,3,\ldots, n_{\alpha} \}$ form complete orthonormal systems (CONS) in the corresponding spaces.
\begin{algorithm}
\textbf{Function SD} 
\BlankLine
\SetKwInOut{Input}{input}
\SetKwInOut{Output}{output}
\Input{$n, n_1, n_2, \mket{\psi}$ where $n=n_1 \cdot n_2, \mket{\psi} \in C^{n_1 \cdot n_2}$}
\Output{$[ r, {\left( \mket{\psi^1_{i}} \right)}_{i=1,2,\ldots, n_1}, {\left( \mket{\psi^2_{i}} \right)}_{i=1,2,\ldots, n_2} ]$}
\BlankLine
\emph{remark:}  $[ r, {\left( \mket{\psi^1_{i}} \right)}_{i=1,2,\ldots, n_1}, {\left( \mket{\psi^2_{i}} \right)}_{i=1,2,\ldots, n_2} ] = SD (\mket{\psi}, (n_1, n_2)) $
\label{lbl:function:sd}
\end{algorithm}

We will also use the following function which computes the complete list of all 2-partitions of a given set $X=\{x_1, x_2, \ldots, x_n\}$.

Let us mention that it is not quite trivial to formulate the algorithm which computes the list of desirable two-partitions of a given n-th element set $X=\{x_1, x_2, \ldots, x_n\}$. However using the methods presented in \cite{Key12} this goal can be obtained.
\begin{algorithm}
\textbf{Function 2p-Par}
\BlankLine
\SetKwInOut{Input}{input}
\SetKwInOut{Output}{output}
\BlankLine
\Input{$X=\{x_1, x_2, \ldots, x_n\}$}
\Output{$2 \wp p$, the list of all 2-partitions of $X$}
\BlankLine
\emph{remark:}  $[ ((X_i,Z_i),i=1,2,\ldots) ] = 2p-Par (X)$
\label{lbl:function:2p:par}
\end{algorithm}
Now we are ready to formulate heuristic version of our algorithm (depicted as Alg.~(\ref{lbl:function:separable:or:entangled:case:of:vector:states})).

\begin{algorithm}
\DontPrintSemicolon
\caption{Separable or Entangled? The case of vector states}
\BlankLine
\SetKwInOut{Input}{input}
\SetKwInOut{Output}{output}
\BlankLine
\Input{$\nH = \nH_1 \otimes \nH_2 \otimes \ldots \otimes \nH_n$, $n = k_1 \cdot k_2 \cdot \ldots k_n$, $\mket{\psi} \in \partial E(H_1 \otimes H_2 \otimes \ldots \otimes \nH_n)$ where $k_i = \dim \nH_i$, $\nH_i \cong C^{k_i}$}
\Output{$\pi \in \wp(n)$ such that $\mket{\psi}$ is $\pi$-separable, in particular  if $\pi = \pi_{\min}$  then $\mket{\psi}$ is completely separable.}
\textbf{Algorithm}
\BlankLine
set $X = I_n = \{1,2,\ldots,n\}; \pi = [ \, ]$ \;
\While{$X| \geq 1$}{
	set $N = |X|$ \;
	\For{$k=1:N$}{
		\If{$|X[k]| = 1$}{
			$X=X \setminus X[k]$ \;
			$\pi=[\pi, X[k]]$ \;
		}
	}
	set $N = |X|$ \;
	\For{k=1:N}{
		2pL[k] = 2p-Par(X[k]) \; 
		set $l=1$ \;
		\While{$l<|2pL[k]|$}{
		\For{$(Y,Z) \in 2pL[k]$}{
			$\left( r_{Y,Z}, \mket{\psi_Y} \mket{\psi_Z} \right) = SD(\mket{\psi_{X[k]}}, (Y,Z))$ \;
			\If{$r_{Y,Z}=1$}{
				set $X[k]=(Y,Z)$ \;
				stop for \;
			}
			\Else{
				go to the next $(Y,Z) \in 2pL[k]$ \;
				$l=l+1$
				}
			}
		}
		\If{$l = |2pL[k]|$}{
			$X = X \setminus X[k]$ \;
			$\pi=[\pi, X[k]]$ \;
		}
	}
}
\label{lbl:function:separable:or:entangled:case:of:vector:states}
\end{algorithm}

\section{A general multipartite system. The case of mixed states} \label{lbl:general:multipartite:system:mixed:states:case}

A general state $\rho \in E(\nH_1 \otimes \nH_2 \otimes \ldots \otimes \nH_n)$ will be called $\pi$-separable state, for some $\pi = (X_{\alpha},\alpha = 1,\ldots,k) \in \wp(n)$ , iff there exists a representation of $\rho$ of the form
\begin{equation}
\rho = \sum_{\alpha} P_{\alpha} \underset{X \in \pi}{\otimes} \rho^{X}_{\alpha},
\end{equation}
where $\rho^{X}_{\alpha} \in E(\nH^{X})$ for each $\alpha$ and moreover $\rho$ is not $\pi^{'}$-separable for any $\pi^{'}$ finer that $\pi$.

There exists plenty of different criterions answering (mostly partially only) the question whether a given state of bipartite system is separable or entangled. Without specifying the concrete form of the criterion used we assume that an appropriate tool for this purpose is available. It will be named as 2p-Oracle.

\begin{algorithm}
\textbf{2p-Oracle function}
\BlankLine
\SetKwInOut{Input}{input}
\SetKwInOut{Output}{output}
\BlankLine
\Input{$\nH=\nH_A \otimes \nH_B$ where $\rho \in E(\nH_A \otimes \nH_B)$}
\Output{YES, if $\rho$ is entangled, NO if $\rho$ is separable}
\BlankLine
\label{lbl:function:2p:oracle}
\end{algorithm}

\begin{remark}
The best known examples of such oracles although working only for a particular classes of states are: PPT-criterion, certain witness construction, see \cite{Key5}, \cite{Key6}, \cite{Key4}. In general the problem of answering the question whether a given quantum state $\rho$ is entangled or not is NP-Hard problem \cite{Key17}.
\end{remark}

Now, we are ready to formulate our algorithm (depicted as Alg.~(\ref{lbl:function:separable:or:entangled:case:of:mixed:states})) by which we can answer the question whether a given general state $\rho \in E(\nH_1 \otimes \nH_2 \otimes \ldots \otimes \nH_n)$ is completely or partially separable and in the second case the localization of the corresponding $\pi \in \wp(n)$ is obtained as well.

\begin{algorithm}
\DontPrintSemicolon
\caption{Separable or Entangled? General n-partite states}
\BlankLine
\SetKwInOut{Input}{input}
\SetKwInOut{Output}{output}
\BlankLine
\Input{$\nH = \nH_1 \otimes \nH_2 \otimes \ldots \otimes \nH_n$, $n = k_1 \cdot k_2 \cdot \ldots k_n$, $\mket{\psi} \in \partial E(H_1 \otimes H_2 \otimes \ldots \otimes \nH_n)$ where $k_i = \dim \nH_i$, $\nH_i \cong C^{k_i}$}
\Output{$\pi \in \wp(n)$ such that $\rho$ is $\pi$-separable}
\textbf{Algorithm}
\BlankLine
set $X = I_n = \{1,2,\ldots,n\}; \pi = [ \, ]$ \;
\While{$X| \geq 1$}{
	set $N = |X|$ \;
	\For{$k=1:N$}{
		\If{$|X[k]| = 1$}{
			$X=X \setminus X[k]$ \;
			$\pi=[\pi, X[k]]$ \;
		}
	}
	set $N = |X|$ \;
	\For{k=1:N}{
		2pL[k] = 2p-Par(X[k]) \; 
		set $l=1$ \;
		\While{$l<|2pL[k]|$}{
			\For{$(Y,Z) \in 2pL[k]$}{
				$\left( r_{Y,Z}, \mket{\psi_Y} \mket{\psi_Z} \right) = SD(\mket{\psi_{X[k]}}, (Y,Z))$ \;
				\If{2p-Oracle($\rho$,$Y$,$Z$)}{
					set $X[k]=(Y,Z)$ \;
					stop for \;
				}
				\Else{
					go to the next $(Y,Z) \in 2pL[k]$ \;
					$l=l+1$
				}
			}
		}
		\If{$l = |2pL[k]|$}{
			$X = X \setminus X[k]$ \;
			$\pi=[\pi, X[k]]$ \;
		}
	}
}
\label{lbl:function:separable:or:entangled:case:of:mixed:states}
\end{algorithm}

\section{Summary} \label{lbl:summary}

A general method for detecting and localizing partial and complete as well entanglement for general multipartite states is presented.

It was shown that the question of detection and localizing (partial) entanglement of quantum states describing multipartite systems can be answered by using exclusively tools worked out for the two-partite case. However the serious drawback of methods introduced here is the non-polynomial computational complexity of our algorithms presented. A rough estimate gives quickly the computational complexity of both algorithms presented as $O(2^n)$. The source of such big complexity is the length of the list of all two-partitions of a given set X.

However, for small quantum registers composed of small number $n$  e.g. 20 -- 25 quantum logical units like qubits or qudits which is the case of all quantum computer simulators available at present (see \cite{Key14}) our methods might be quite useful.

The methods introduced in the present contribution could find straightforward applications to several problems of quantum information processing tasks. Let us mention two of them:
\begin{itemize}
\item in the process of tracing of entanglement when executing quantum programs on Quantum Computer Simulators,
\item in the search for a new effectively simmulable quantum circuits by using the methods of \cite{Key15}.
\end{itemize}

A natural semiorder relation with respect to the amount of entanglement included is that induced by the action of (S)LOCC-class of operations. In the case of two-partite systems this notion seems to be quite well understood \cite{Key4}, \cite{Key5}, \cite{Key6} although several basic question are still to be resolved even in this case. However, in the case of multipartite systems several problems arise at the very beginning. For example, let us consider two quantum states $\rho_1$, $\rho_2$ of a composite nth-partite system and let us assume that $\rho_1$ is $\pi_1$-entangled and $\rho_2$ is $\pi_2$-entangled where of course $\pi_i = \left( X^i_1, X^i_2, \ldots, X^i_{k_i} \right), i=1,2$ are the corresponding partitions of the set of indices $\{1, 2, 3, \ldots, n\}$.

From the very definitions of (S)LOCC class of operation we must restrict ourselves to the situation $\pi_1 \prec_{f} \pi_2$ or $\pi_2 \prec_f \pi_1$ in order to define the local actions on both states $\rho_1$ and $\rho_2$. So, let us assume that $\pi_1 = (X_1, X_2, \ldots, X_k) \prec \pi_2 = (Y_1, Y_2, \ldots, Y_l)$. By (S)LOCC-class of operations we mean local with respect to the finer decomposition $(X_1, X_2, \ldots, X_k)$ completely positive operators together with general but local measurement equipments (supported with classical communications channels) \cite{Key18}. We say that $\rho_1 \underset{S(LOCC)}{\prec} \rho_2$ iff there exists (S)LOCC-class operations that transforms $\rho_2$ into the state $\rho_1$. From the above definitions it follows that we should expect much linear chains in $POSET \, (E(\nH_1 \otimes \nH_2 \otimes \ldots \otimes \nH_n),\underset{S(LOCC)}{\prec})$  then in the bipartite case. In particular it follows from \cite{Key16} that there exists many locally maximal states with respect to this semi-order.

\end{document}